\documentclass[a4paper,11pt]{article}
\usepackage{pos}

\title{Geometric scaling in elastic $pp$ collisions }

\author*{Michał Praszałowicz}

\affiliation{Institute of Theoretical Physics, Jagiellonian University,\\
  Łojasiewicza 11, Kraków, Poland}

\emailAdd{michal.praszalowicz@uj.edu.pl}

\abstract{Geometric scaling was conjectured and observed at the ISR more than 50 years ago. 
We argue that it still holds at the LHC. We show that the dip-bump structures of the differential elastic cross-sections 
exhibit a remarkable regularity, namely that the ratio of the bump to dip positions is constant from 23 GeV to 13 TeV. 
Using crossing symmetry, analicity and the optical theorem, we identify the imaginary and real parts of the scattering amplitude. 
This allows us to compute the  $\rho$
parameter and the ratio of the bump to dip values of the differential 
$pp$ cross-section. Finally, we discuss the energy dependence of the total elastic cross-section as compared to the total cross-section, 
and the violation of geometrical scaling outside the dip-bump region at the LHC.}

\FullConference{Proceedings of the Corfu Summer Institute 2025 "School and Workshops on Elementary Particle Physics and Gravity" (CORFU2025)\\
27 April - 28 September, 2025\\
Corfu, Greece\\}

\begin{document}
\maketitle

\section{Introduction}
\label{sec:intro}
In 1973, Dias de Deus \cite{DiasDeDeus:1973lde} conjectured that  $pp$ cross-sections (elastic, total and inelastic) exhibit so-called geometric scaling (GS). The GS hypothesis
states that  the elastic scattering amplitude, which formally depends on two variables $s$ and $t$ (or equivalently $b$, {\em i.e.} the impact parameter), depends in fact only on the combination
$t R^2(s)$ (or $b/R(s)$), where $R$ is an energy-dependent parameter called the {\em interaction radius}. In the impact parameter space
\begin{equation}
\sigma_{\text{tot}}(s)    =2%
{\displaystyle\int}
d^{2}\boldsymbol{b}\,\operatorname{Im}T_{\text{el}}(s,b), ~~~
\sigma_{\text{el}}(s)    =%
{\displaystyle\int}
d^{2}\boldsymbol{b}\,\left\vert T_{\text{el}}(s,b)\right\vert ^{2}
\label{eq:sigmas}%
\end{equation}
and the inelastic cross-section is given as the difference $\sigma_{\text{tot}}(s) -\sigma_{\text{el}}(s)$. 
Note that in this formulation the elastic amplitude is dimensionless. If $T_{\rm el}(s,b)=T_{\rm el}\big(B=b/R(s)\big)$ then
\begin{equation}
\sigma_{\text{tot}}(s)    =2 R^2(s)%
{\displaystyle\int}
d^{2}\boldsymbol{B}\,\operatorname{Im}T_{\text{el}}(B), ~~~
\sigma_{\text{el}}(s)    = R^2(s)%
{\displaystyle\int}
d^{2}\boldsymbol{B}\,\left\vert T_{\text{el}}(B)\right\vert ^{2}
\label{eq:sigmasb}%
\end{equation}
and all integrated $pp$ cross-sections should exhibit the same
energy behavior, provided that both real and imaginary parts of the amplitude scale in the same way, or
the real part, which is known to be small, can be neglected. 
Such a uniform energy dependence has indeed been observed at the ISR. To illustrate this behavior we have fitted~\cite{Baldenegro:2024vgg} power law energy dependence
separately for the ISR and for the LHC
\begin{equation}
\sigma_{\rm tot}^{\rm ISR}\sim \sigma_{\rm el}^{\rm ISR}\sim W^{0.11},~~~
\sigma_{\rm tot}^{\rm LHC}\sim W^{0.17},~~~ \sigma_{\rm el}^{\rm LHC}\sim W^{0.23}\, ,
\label{eq:Wpower}
\end{equation}
where $W=\sqrt{s}$. The exact value of the exponent for the total cross-section at the LHC reads $\beta=0.1729\pm0.0163$ \cite{Baldenegro:2024vgg}. The different exponents of $W$ for the LHC fits suggest that  GS does not occur at TeV energies.
In Refs.~\cite{Baldenegro:2024vgg,Praszalowicz:2025twy} and \cite{Praszalowicz:2025vdb} we have argued that this is not the case.
The GS still holds at the LHC, however, in the restricted $t$ range.  Here we summarize the results obtained in Refs.~\cite{Baldenegro:2024vgg,Praszalowicz:2025twy}, to which we refer the reader for more details and a complete
list of references.

\section{Dips and bumps}

In momentum space we can rewrite the cross-sections (\ref{eq:sigmas}) as
\begin{equation}
s \sigma_{\text{tot}}(s)    =2\operatorname{Im}\tilde{T}_{\text{el}}(s,0),
~~~~
\sigma_{\text{el}}(s)    =\frac{1}{4\pi s^{2}}%
{\displaystyle\int}
dt\left\vert \tilde{T}_{\text{el}}(s,t)\right\vert ^{2} \, ,
\label{eq:totel}
\end{equation}
where $\tilde{T}_{\rm el}(s,t)$ is a Fourier transform of $T_{\text{el}}(s,b)$ in a normalization where 
 $\tilde{T}_{\rm el}(s,t)$ is dimensionless:
 \begin{equation}
\tilde{T}_{\text{el}}(s,t)=\pi s {\displaystyle\int}_{\! 0}^{\infty}db^2 T_{\text{el}}(s,b) J_0(bq).
\label{eq:tildeTt}
 \end{equation}
Here $q=\sqrt{-t}$. In the first equation (\ref{eq:totel}) we recognize the optical theorem.

\begin{figure}[h]
\centering
\includegraphics[height=4.5cm]{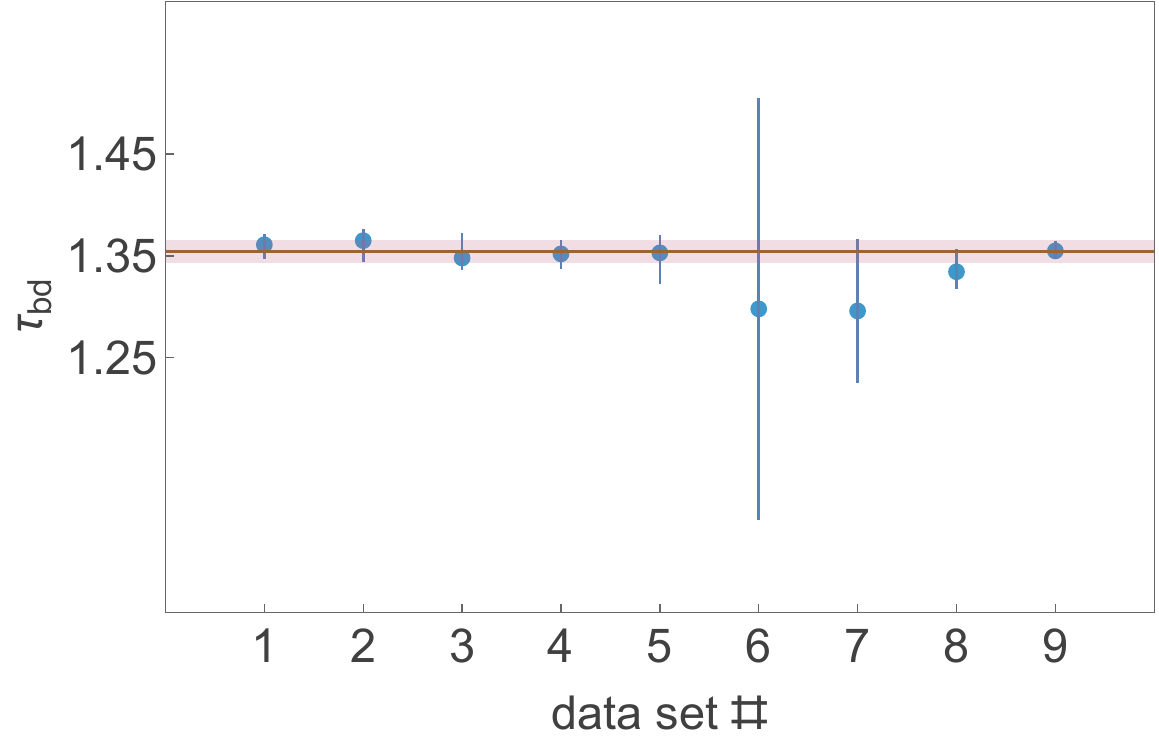}
\caption{Ratio $\mathcal{T}_{\rm bd}=|t_{\rm b}|/|t_{\rm d}|$ for different energies as a function of the data set number from Ref.~\cite{Baldenegro:2024vgg} (
ISR: (1) -- 23, (2) -- 30, (3) -- 45, (4) -- 53, (5) -- 63~GeV,  LHC: (6) -- 2.76, (7) -- 7, (8) -- 8, (9) -- 13~TeV).}%
\label{fig:ratio}%
\end{figure}

Typically, $T_{\text{el}}(s,b)$ is a smooth Gaussian-, Fermi- or step-like function of $b$ \cite{Kamal1975}. Hence $\tilde{T}_{\text{el}}(s,t)$ exhibits
a well-known dip and bump structure governed by the oscillations and zeros of the Bessel function $J_0(b q)$. While the cross-section ratio 
\begin{equation}
{\cal R}_{\rm bd}=(d\sigma_{\rm el}/dt|_{\rm bump})/(d\sigma_{\rm el}/dt|_{\rm dip})
\label{eq:Rbddef}
\end{equation}
is strongly energy dependent at the ISR and saturates~\cite{TOTEM:2020zzr} at $\sim 1.8$ at the LHC (see Fig.~\ref{fig:total}), the ratio of the
 bump to dip {\em postions} 
\begin{equation}
{\cal T}_{\rm bd}=t_{\rm bump}/t_{\rm dip}
\label{eq:Tratio}
\end{equation}
is constant over the wide energy range, from 23~GeV (ISR) to 13~TeV (LHC): 
${\cal T}_{\rm bd}=1.355\pm0.011$ \cite{Baldenegro:2024vgg}. This largely unexpected property, shown in Fig.~\ref{fig:ratio}, 
means that  dips and bumps have a universal
energy dependence $|t_{\rm dip,\, bump}|=\tau_{\rm dip,\, bump}/R^2(s)$, where  $R^2(s)$ cancels
in the ratio (\ref{eq:Tratio}). 

The constancy of ${\cal T}_{\rm bd}$ is a strong indication of GS even at the LHC. Let's explore it assuming for a while that the elastic scattering amplitude $\tilde{T}_{\text{el}}(s,t)$ 
is purely imaginary\,\footnote{This cannot be true everywhere in $t$, since (\ref{eq:tildeTt})  implies that $\tilde{T}_{\text{el}}(s,t)$ has zeros and the cross-section vanishes.} and 
depends on the scaling variable $\tau=-t/R^2(s)$
\begin{equation}
\tilde{T}_{\text{el}}(s,t)=i \operatorname{Im} \tilde{T}_{\text{el}}(\tau)=i\, sR^2(s) \Phi(\tau).
\end{equation}
With this normalization $R^2(s)=\sigma_{\rm tot}(s)/(2\Phi(0))$. To compute $\Phi(\tau)$ let's apply the GS hypothesis to (\ref{eq:tildeTt}) (assuming for convenience that the amplitude
in the impact parameter representation is $i T_{\text{el}}(s,b)$)
\begin{eqnarray}
 \tilde{T}_{\text{el}}(s,t)&=&i \pi s {\int}_{\! 0}^{\infty} db^2 T_{\text{el}}(b/R(s)) J_0(bq) \notag \\
                                    &=&i s R^2(s)\,\pi {\int}_{\! 0}^{\infty} dB^2 T_{\text{el}}(B) J_0(B\sqrt{\tau}),
\label{eq:tildeTtau}
 \end{eqnarray}
which gives
\begin{equation}
\Phi(\tau)=\pi {\int}_{\! 0}^{\infty} dB^2 T_{\text{el}}(B) J_0(B \sqrt{\tau}) .
\label{eq:Phidef}
\end{equation}

\begin{figure}[h]
\centering
\includegraphics[height=6.0cm]{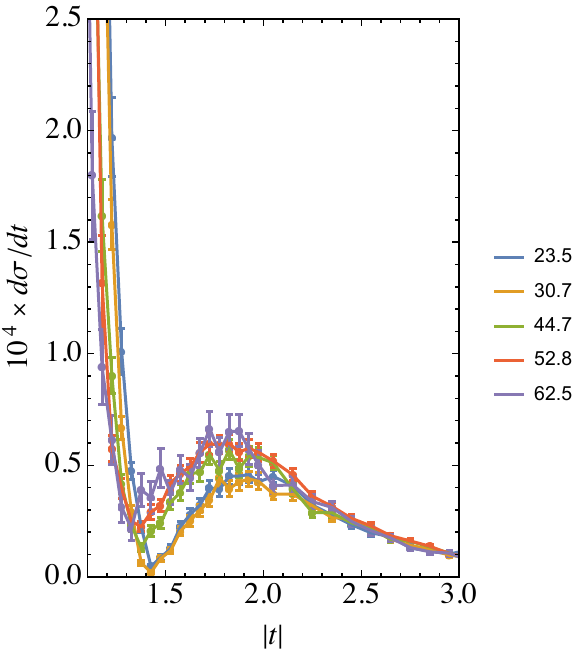}
~\includegraphics[height=6.0cm]{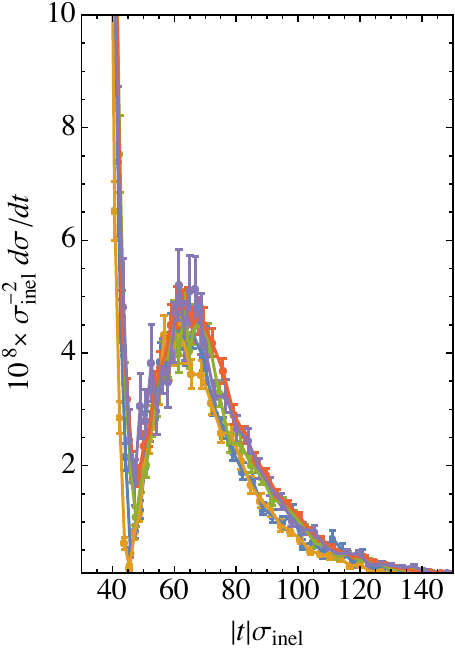}
\caption{Elastic $pp$ cross-section
$d\sigma_{\mathrm{el}}/dt$~[mb/GeV$^{2}$] at the ISR energies in terms of
$|t|$~[GeV$^{2}$] -- left, and the same scaled according to
Eq.~(\ref{eq:Phiscaling}) -- right. We have used, following \cite{Buras:1973km}, $R^2=\sigma_{\rm inel}$ rather than $R^2=\sigma_{\rm tot}$.
This does not matter for the ISR since all integrated cross-sections have the same energy dependence (\ref{eq:Wpower}).}%
\label{fig:isr}%
\end{figure}

Finally, following (\ref{eq:totel}), (\ref{eq:tildeTtau}) and (\ref{eq:Phidef}), the differential elastic cross-section reads
\begin{align}
\frac{d\sigma_{\text{el}}}{dt}  &    = \frac{1}{4\pi}{ R^{4}(s)} \Phi^{2}(\tau) ,
\label{eq:dsigmadt}%
\end{align}
which means that
\begin{equation}
\Phi^{2}(\tau) \sim \frac{1}{\sigma^2_{\rm tot}(s)} \frac{d\sigma_{\text{el}}(s,t)}{dt}
\label{eq:Phiscaling}
\end{equation}
is a universal, energy independent function of the scaling variable $\tau=-t\sigma_{\rm tot}(s)$. This is the essence of geometric scaling.\footnote{Geometric scaling of the $pp$
elastic cross-section should not be confused with geometric scaling of DIS structure functions \cite{Stasto:2000er},  even though the name was borrowed from the $pp$ case.}  
In Ref.~\cite{Buras:1973km}
Buras and Dias de Deus demonstrated, based on the ISR data, that GS really works, except for the dip region \cite{DiasdeDeus:1975ybq}. In Ref.~\cite{Baldenegro:2024vgg} we have repeated their
analysis and the results are shown in Fig.~\ref{fig:isr}.  Note that a single function $R^2(s)=\sigma_{\rm tot}(s)$ handles both the alignment  of dip and bump positions, as well as
the cross-section values. While the first property holds also at the LHC, due to the energy independence of ${\cal T}_{\rm bd}$, the cross-section values can be approximately superimposed by a different function of $s$ \cite{Baldenegro:2024vgg}.

\section{Crossing and the real part of the amplitude}

It is clear that in the vicinity of the dip related to the zero of the Bessel function $J_0(bq)$, one should take into account
the real part of the amplitude, no matter how small it is. To this end we shall observe that in the high energy limit  elastic (complex) amplitude must satisfy the following crossing relation%
\begin{equation}
\tilde{T}_{\text{el}}(u,t)\simeq\tilde{T}_{\text{el}}(-s,t)=\tilde
{T}_{\text{el}}^{\ast}(s,t). \label{eq:crossing}%
\end{equation}
Therefore, following \cite{DiasdeDeus:1975ybq,DiasdeDeus:1977af} we assume%
\begin{equation}
\tilde{T}_{\text{el}}(s,\tau)=isR^{2}(-is)\Phi\left[  \left\vert
t\right\vert R^{2}(-is)\right]  ,
\label{eq:ampcross}
\end{equation}
which satisfies (\ref{eq:crossing}).

To identify real and imaginary part of the amplitude we use the following trick \cite{DiasdeDeus:1975ybq,DiasdeDeus:1977af}
\begin{equation}
-is=e^{y-i\pi/2}
\end{equation}
and expand (\ref{eq:ampcross}) around $y$. First we expand $R^2(-is)$
\begin{equation}
R^{2}(-is) \rightarrow R^{2}\left(  y-i\frac{\pi}{2}\right)  \simeq
R^{2}(y)-i\frac{\pi}{2}\frac{dR^{2}(y)}{dy} \, .\label{eq:expansionR}%
\end{equation}
Next, we expand $\Phi$%
\begin{equation}
\Phi\Big( \left\vert t\right\vert R^{2}(-is)\Big) \simeq\Phi\left(
\tau\right)  -i\frac{\pi}{2}\frac{d\Phi\left(  \tau\right) }{d\tau}%
\frac{dR^{2}(y)}{dy}\left\vert t\right\vert . \label{eq:expansion}%
\end{equation}
Keeping only linear terms from expansions (\ref{eq:expansionR}) and
(\ref{eq:expansion}) we get~\cite{Praszalowicz:2025twy,DiasdeDeus:1975ybq,DiasdeDeus:1977af}%
\begin{align}
\operatorname{Im}\tilde{T}_{\text{el}}(s,\tau)  &  = sR^{2}(y)\Phi\left(
\tau\right) ,\nonumber\\
\operatorname{Re}\tilde{T}_{\text{el}}(s,\tau) &  =s\frac{\pi}{2}\frac
{dR^{2}(y)}{dy}\frac{d}{d\tau}\left(  \tau\Phi\left( \tau\right)  \right) .
\label{eq:ImRe}%
\end{align}
Expanding further up to quadratic terms one has two additional contributions, but only to the imaginary part
\begin{equation}
\operatorname{Im}\tilde{T}_{\text{el}}(s,\tau)   =sR^{2}\left(  \Phi
-\frac{\pi^{2}}{8}\left[  \frac{1}{R^{2}}\frac{d^{2}R^{2}}{dy^{2}}\right]
\frac{d}{d\tau}(\tau\Phi)-\frac{\pi^{2}}{8}\left[  \frac{1}{R^{2}}\frac
{dR^{2}}{dy}\right] ^{2} \frac{d}{d\tau}\Big(\tau^{2} \frac{d}{d\tau}\Phi \Big) \right).
\end{equation}
We have checked that these extra terms are numerically small, and therefore in the following we shall use (\ref{eq:ImRe}).

\begin{figure}[h]
\centering
\includegraphics[height=3.5cm]{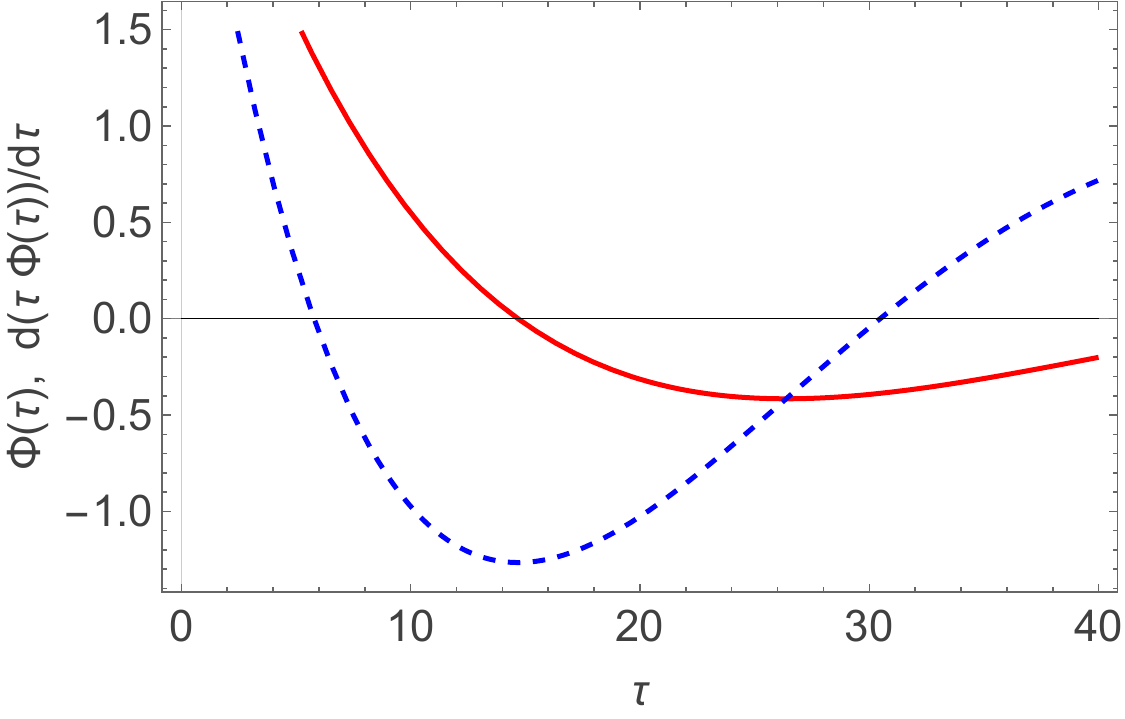} ~~ \includegraphics[height=3.5cm]{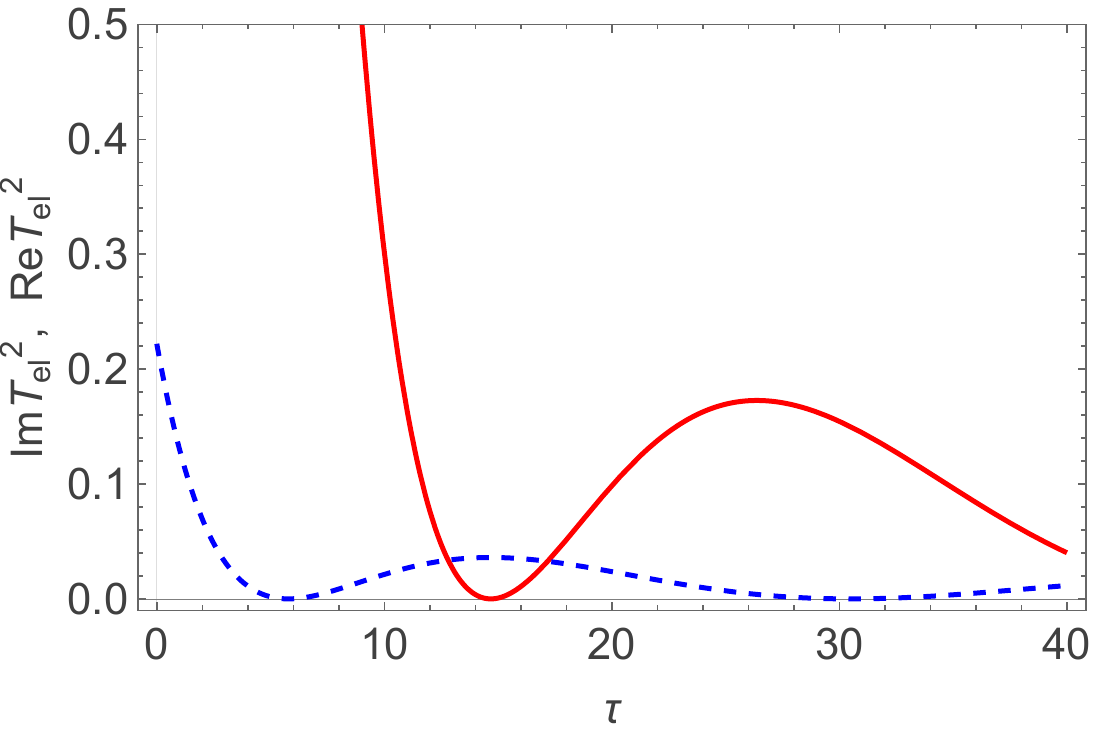}
\vspace{-0.2cm} \caption{Left panel: function $\Phi(\tau)$ from Eq.~(\ref{eq:PhiHD2}) -- solid red line,
and $d(\tau \Phi(\tau))/d\tau$ -- dashed blue. Right panel: Contributions of imaginary (solid red)
and real (dashed blue) parts of the scattering amplitude to the elastic cross section. For the real
part the $\rho$ parameter is 0.15. }%
\label{fig:Phi}%
\end{figure}

In order to understand the interplay between real and imaginary part of the amplitude it is instructive to consider a black disk toy model 
where~\footnote{Remember we have already factored out $i$ from the amplitude, see (\ref{eq:tildeTtau}).}  $T_{\mathrm{el}}(B) =\Theta(1-B)$.
From (\ref{eq:Phidef}) we get \cite{Praszalowicz:2025twy}
\begin{equation}
\Phi(\tau)={2 \pi}\frac{J_{1}(\sqrt{\tau})}{\sqrt{\tau}},
\label{eq:PhiHD2}
\end{equation}
which is plotted in the left panel of Fig.~\ref{fig:Phi} as a solid red line. In the right panel we plot $\Phi^2$,
which is responsible for the imaginary part contribution to the cross section.
In the same figure we also plot as a blue dashed line $d(\tau \Phi(\tau))/d\tau$ corresponding to the real part of the amplitude  (\ref{eq:ImRe})  and
-- in the right panel  --  $\rho^2 (d(\tau \Phi(\tau))/d\tau)^2$. Here $\rho$ corresponds to the relative strength of the real part contribution to the cross-section
(for illustration purposes we have assumed $\rho=0.15$).

One can clearly see from Fig.~\ref{fig:Phi} that the cross-section dip corresponds to the zero of $\Phi$, and the bump to its minimum. Therefore we have
\begin{equation}
\Phi(\tau_{\mathrm{dip}})=0,~~\frac{d}{d\tau}\Phi(\tau)|_{\mathrm{bump}%
}=0.
\label{eq:dipbumpconds}%
\end{equation}
As seen from the right panel of Fig.~\ref{fig:Phi}, the value of the cross-section at the dip is entirely given by the real part. 
Outside the dip the real part can be safely neglected.
This pattern will hold for more realistic profiles as well.

\section{Phenomenology}

There are three immediate predictions, which follow from the discussion in the previous section. The first one
is an absolute prediction for the  parameter  $\rho={\rm Re}T_{\rm el}(s,t=0)/{\rm Im}T_{\rm el}(s,t=0)$ ~\cite{Praszalowicz:2025twy} 
\begin{equation}
\rho = \frac{\pi}{2 }\frac{1}{R^{2}(y)}%
\frac{dR^{2}(y)}{dy} \, ,
\label{eq:rhopred}%
\end{equation}
which is governed entirely by the energy dependence of $R^2(y)$.

Next, we have a prediction for ${\cal R}_{\rm bd}$ (\ref{eq:Rbddef}). To this end we have to compute the elastic $pp$ cross-section at the dip and at the bump \cite{Praszalowicz:2025twy}. We obtain
\begin{align}
\Phi(\tau_{\mathrm{dip}})=0 \rightarrow &  \operatorname{Im}\tilde
{T}_{\text{el}}(s,\tau_{\mathrm{dip}})=0,
~~~
 \operatorname{Re}\tilde{T}_{\text{el}}(s,\tau_{\mathrm{dip}})=s\frac{\pi
}{2}\frac{dR^{2}(y)}{dy}\frac{d}{d\tau} \Phi(\tau_{\mathrm{dip}}),
\\
\frac{d}{d\tau} \Phi(\tau_{\mathrm{bump}}) =0 \rightarrow &  \operatorname{Im}%
\tilde{T}_{\text{el}}(s,\tau_{\mathrm{bump}})= sR^{2}(y)\Phi(\tau
_{\mathrm{bump}}),
~~~
\operatorname{Re}\tilde{T}_{\text{el}}(s,\tau_{\mathrm{bump}})=s\frac{\pi
}{2}\frac{dR^{2}(y)}{dy} \Phi(\tau_{\mathrm{bump}}) \, . \nonumber
\end{align}
Therefore%
\begin{align}
\left.  \frac{d\sigma}{dt}\right\vert _{\text{dip}}  &  
=\frac{R^{4}(y)}{4\pi} \, \rho^{2}(y) \, \left(  \tau_{\text{dip}}\frac{d}{d\tau}
\Phi (\tau_{\mathrm{dip}})\right)^{2},%
~~~~
\left.  \frac{d\sigma}{dt}\right\vert _{\text{bump}}  &  
 =\frac{R^{4}(y)}{4\pi} \, \Big(  1+\rho^{2}\left(  y\right)  \Big)\,
\Phi^{2}(\tau_{\mathrm{bump}}) \, .
\end{align}
Hence, we have a one parameter prediction for the ratio
\begin{equation}
{\cal R}_{\rm bd}(s)=\frac{d\sigma/dt(t_{\text{bump}})}{d\sigma/dt(t_{\text{dip}})}=c_{0}%
\frac{1+\rho^{2}\left(  y\right)  }{\rho^{2}\left(  y\right)  },
\label{eq:ratioRbd}
\end{equation}
where the constant $c_0$ is given in terms of the  function $\Phi$%
\begin{equation}
c_{0}=\frac{\Phi^{2}(\tau_{\mathrm{bump}})}{\left(  \tau_{\text{dip}}\frac
{d}{d\tau}\Phi (\tau_{\mathrm{dip}}) \right)  ^{2}} \, 
\label{eq:Rbdc0}
\end{equation}
and we will treat it as a free parameter.

Finally, equations (\ref{eq:ImRe}) allow to predict the energy behavior of the total elastic cross section~\cite{Praszalowicz:2025twy}
\begin{align}
\sigma_{\text{el}}(s)  & 
 =\frac{1}{4\pi R^{2}(y)}\left[  {R^{4}(y)}%
{\int}d\tau\, {\Phi}^{2}{(\tau)+}\left(  {\frac{\pi}{2}\frac{dR^{2}(y)}{dy}}\right)^{2}%
{\int}d\tau\left(  {\frac{d}{d\tau}\left(  \tau\Phi (\tau)\right)  }\right)^{2}\right] \notag \\
 &=\frac{R^{2}(y)}{4\pi}\left(  1+c_{1}\rho^{2}(y)\right)  \times%
{\displaystyle\int}
d\tau\, {\Phi}^{2}{(\tau)}\, ,%
\label{eq:elvstot}
\end{align}
where%
\begin{equation}
c_{1}=\frac{%
{\displaystyle\int}
d\tau\left(  {\frac{d}{d\tau}\left(  \tau\Phi (\tau)\right)  }\right)
^{2}}{%
{\displaystyle\int}
d\tau{\Phi}^{2}{(\tau)}}.
\label{eq:sigeltot}
\end{equation}
We see that the energy dependence of $\sigma_{\text{el}}(s)$ is modified with respect to $\sigma_{\text{tot}}(s)$ by a factor $1+c_1\rho^2(s)$.

\begin{figure}[h]
\centering
\includegraphics[width=6cm]{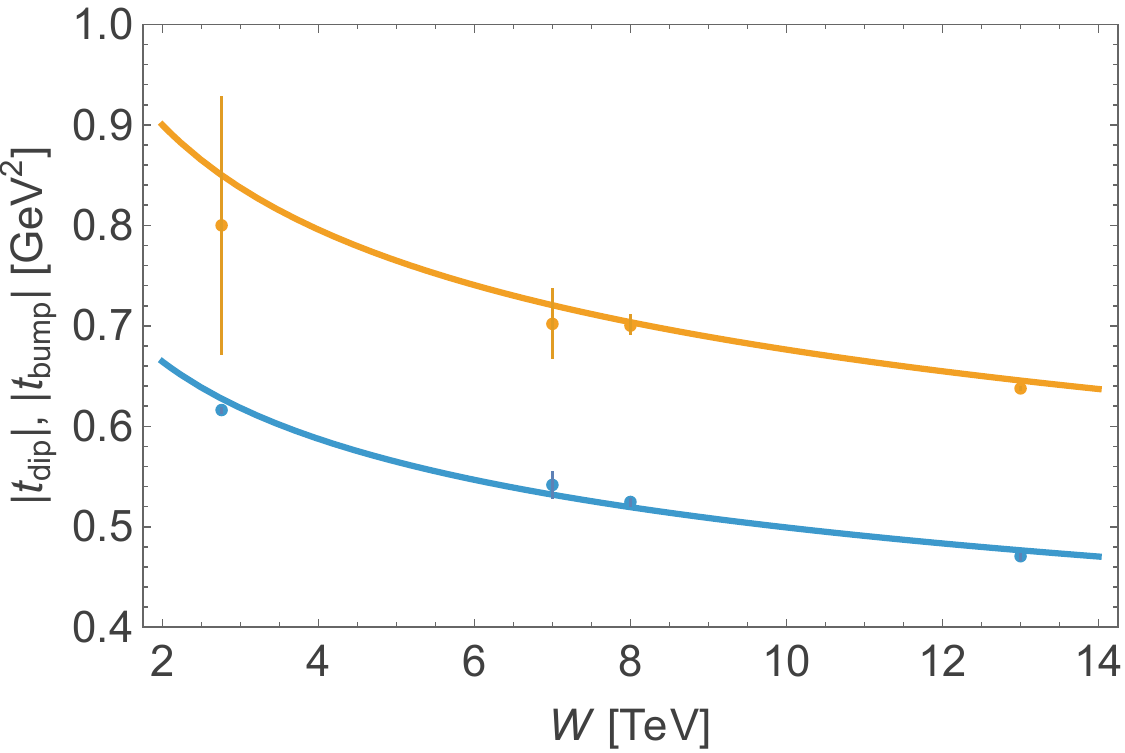} 
\vspace{-0.2cm}\caption{Fit to dip
and bump positions of Eq.~(\ref{eq:dipbumpfits}) with $\beta=0.1686$ at LHC energies.}%
\label{fig:dipfit}%
\end{figure}

To verify the above predictions based on experimental data, we need to know two functions $R^2(s)$ and in principle also  $\Phi(\tau)$. Note that $R^2$ can be found from
two independent sources, namely from  the energy dependence of the total cross-section, see (\ref{eq:sigmasb}) and (\ref{eq:Wpower}), and from the energy dependence 
of  $t_{\rm dip,bump}$. In Ref.~\cite{Baldenegro:2024vgg} we have shown that both dip and bump positions at the LHC energies can be fitted
with a power law function, {\em i.e.} $t_{\rm dip,bump}(W)=B_{\rm dip,bump} W^{-\beta}$. However,
due to large uncertainties on the bump positions (see Fig.~\ref{fig:dipfit}), we have performed
only a fit to the dips and for the bumps we have used ${\cal T}_{\rm bd}=1.355$:
\begin{equation}
t_{\mathrm{dip}}(W)=(0.732\pm 0.003)\times(W/(1~\mathrm{TeV}))^{-0.1686\pm 0.0027}
~~~{\rm and}~~
t_{\mathrm{bump}}(W)=1.355 \times t_{\mathrm{dip}}(W) \, .
\label{eq:dipbumpfits}%
\end{equation}
We have checked that $t_{\mathrm{bump}}(W)$ of Eq.~(\ref{eq:dipbumpfits})
gives  $\chi^2\approx 1$. The fits are shown in Fig.\ref{fig:dipfit}. We see that, within experimental accuracy,
both $\sigma_{\rm tot}$ and $t_{\rm dip, bump}$ have the same energy dependence. Therefore in the following we shall
use $R^2(s)=\sigma_{\rm tot}(s)$ shifting the normalization factor $2\Phi(0)$ to the definition of $\Phi$. This change of normalization
does not affect the ratios in (\ref{eq:rhopred}), (\ref{eq:ratioRbd}) and (\ref{eq:sigeltot}).

\begin{figure}[h]
\centering
\includegraphics[width=6cm]{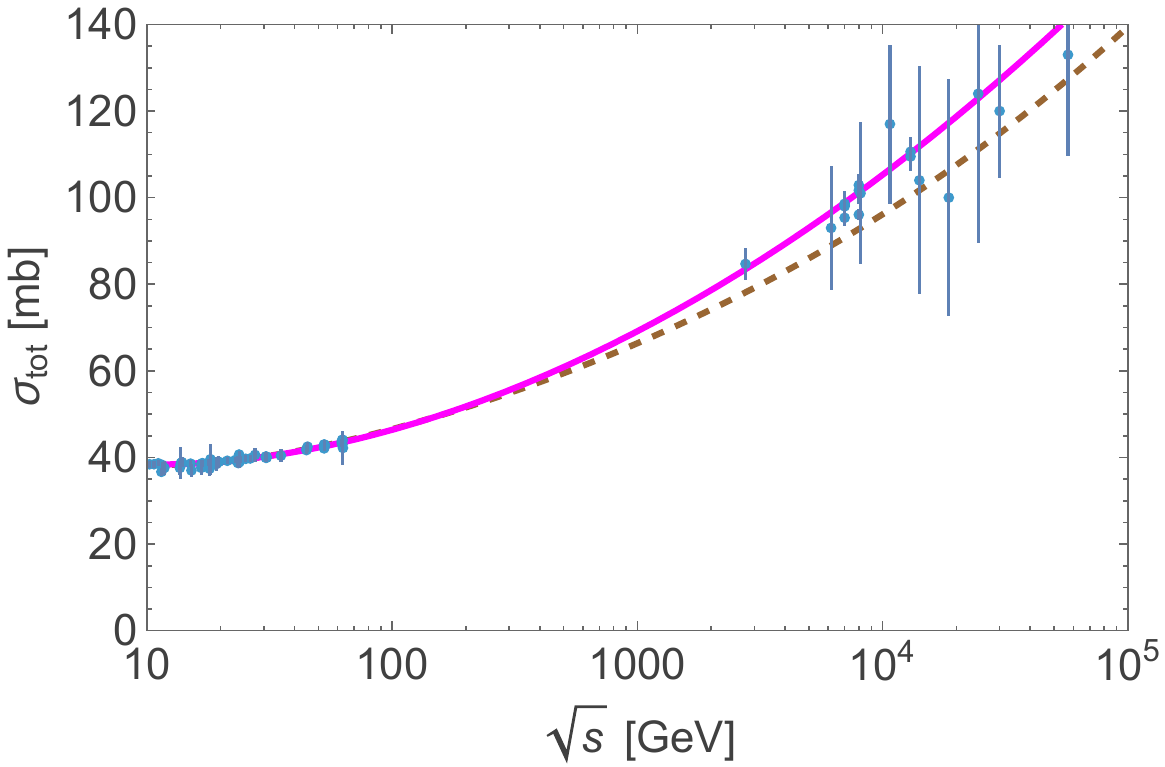} \includegraphics[width=6cm]{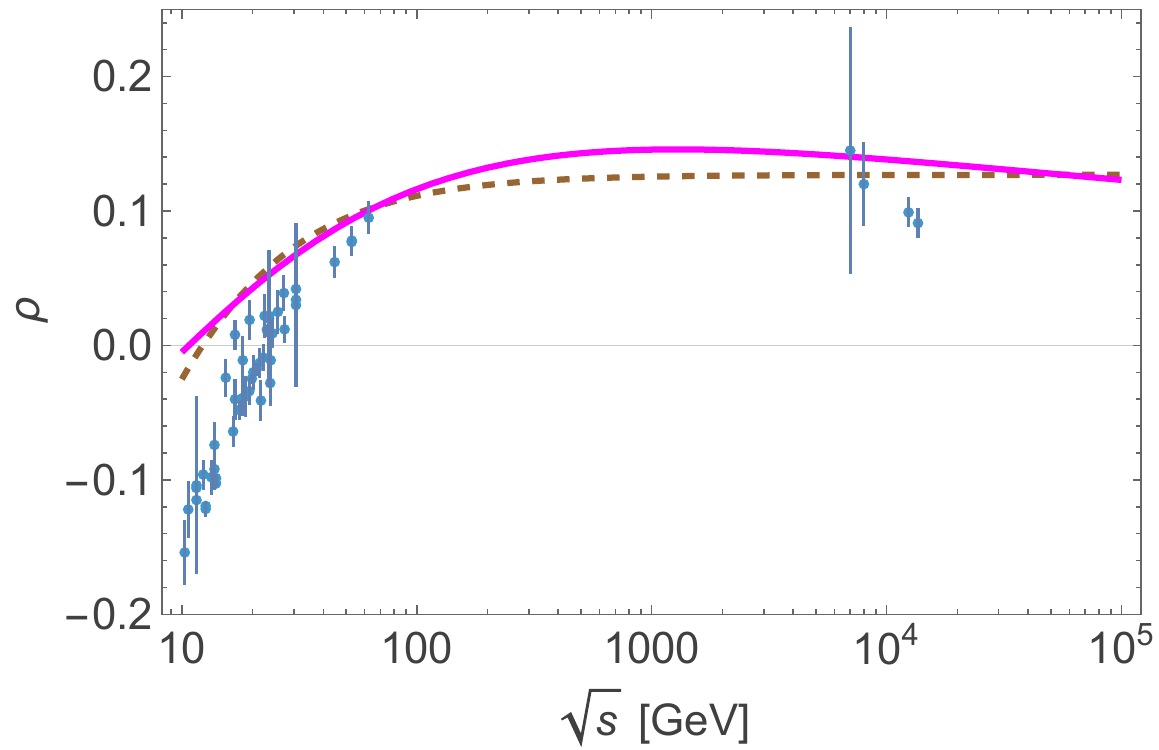} 
\vspace{-0.2cm} \caption{Left: total $pp$
cross-section in mb as a function of $\sqrt{s}$ in GeV. Points at small
$\sqrt{s} < 100$~GeV are from the ISR, points above 1~TeV are from the LHC
(small error bars) and from cosmic rays. Right: Parameter $\rho$ (\ref{eq:rhopred})
as a function of $\sqrt{s}$ in GeV. 
Solid magenta line corresponds to the
PDG parametrization (\ref{eq:parPDG}) and brown dashed line to (\ref{eq:parDL}%
). Data from \cite{PDG2022} and also from \cite{TOTEM:2017sdy}.}%
\label{fig:total}%
\end{figure}

In order to compare (\ref{eq:rhopred}), (\ref{eq:ratioRbd}) and (\ref{eq:elvstot}) with the data we used two analytic
parametrizations of the total $pp$ cross-section data. The first one is the COMPETE parametrization
\cite{Cudell:2001pn} quoted in PDG~2010~\cite{PDG2010}
\begin{equation}
\sigma_{\mathrm{tot}}^{\mathrm{PDG}}(s)=Z+C \ln^{2}\left(  \frac{s}{s_{0}%
}\right)  +Y_{1}\left(  \frac{s}{s_{1}}\right)  ^{-\eta_{1}}-Y_{2}\left(
\frac{s}{s_{1}}\right)  ^{-\eta_{2}} \, ,
\label{eq:parPDG}%
\end{equation}
where $Z=35.45$~mb, $C=0.308$~mb, $Y_{1}=42.53$~mb, $Y_{2}=33.34$~mb,
$s_{0}=28.94$~GeV$^{2}$, $s_{1}=1$~GeV$^{2}$ and $\eta_{1}=0.458$, $\eta
_{2}=0.545$. The second one is from Ref.~\cite{Donnachie:1992ny} by Donnachie
and Landshoff
\begin{equation}
\sigma_{\mathrm{tot}}^{\mathrm{DL}}(s)= A \left(  \frac{s}{s_{1}}\right)
^{\alpha}+B\left(  \frac{s}{s_{1}}\right)  ^{\beta} \ ,
\label{eq:parDL}%
\end{equation}
where $A=21.70$~mb, $B=56.08$~mb, $\alpha=0.0808$, $\beta=-0.4525$ and
$s_{1}=1$~GeV$^{2}$. They are plotted in the left panel Fig.~\ref{fig:total} together with
data points (see the web page of PDG~2022~\cite{PDG2022}).

In the right panel of Fig.~\ref{fig:total} we plot parameter $\rho$ computed according
to (\ref{eq:rhopred}). Note that (\ref{eq:rhopred}) is a parameter free prediction of GS and
depends only on the energy dependence of $\sigma_{\rm tot}$.
We see that both low and high energy data
are well reproduced, both normalization and energy dependence. The last two points
in the right panel of   Fig.~\ref{fig:total} correspond to
two different estimates of $\rho$ by TOTEM~\cite{TOTEM:2017sdy}. Their rapid
decrease with energy  has been attributed the odderon \cite{TOTEM:2020zzr}.

\begin{figure}[h]
\centering
\includegraphics[width=6.cm]{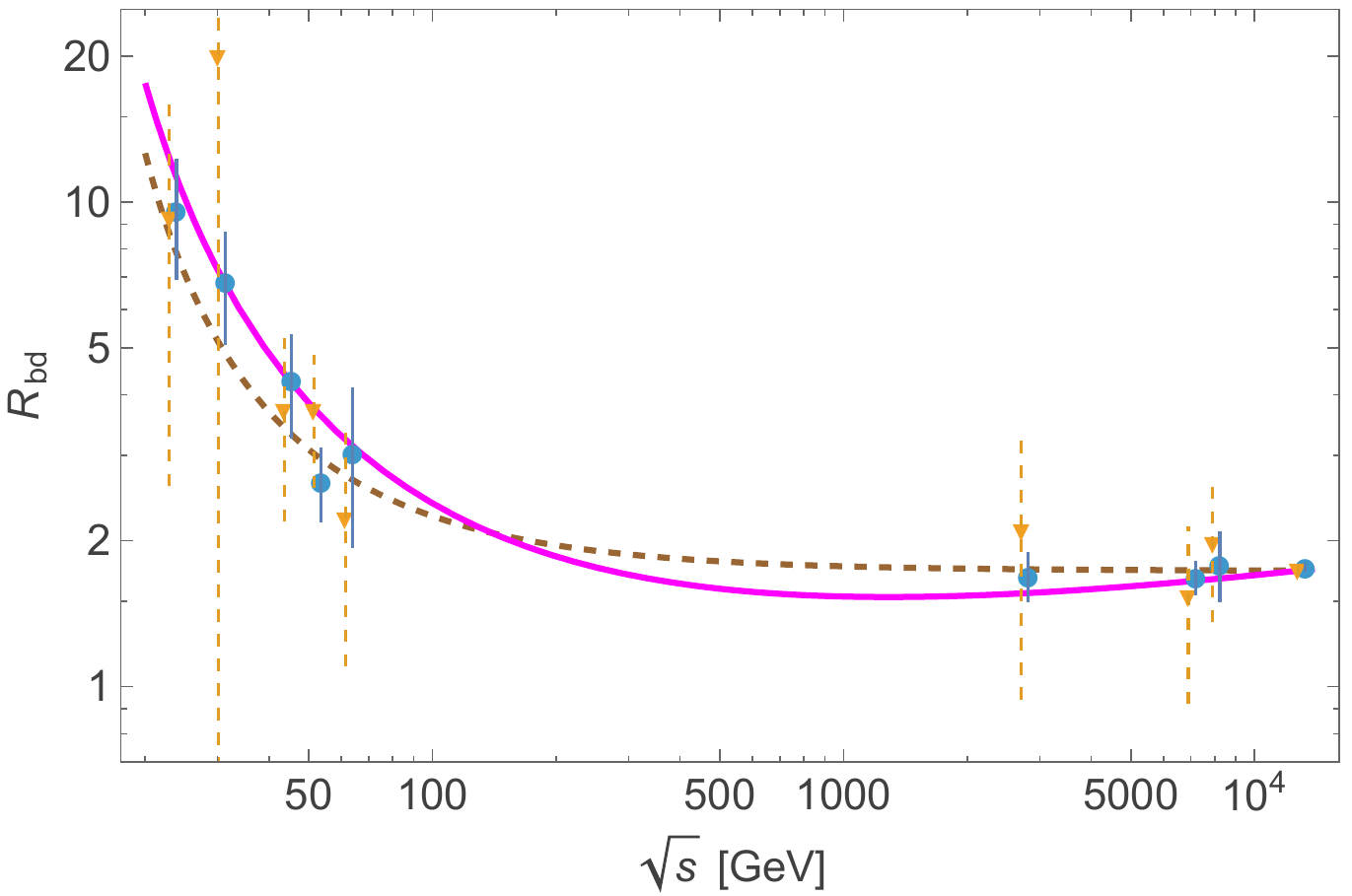} 
\vspace{-0.2cm}\caption{Ratio
${\cal R}_{\mathrm{bd}}$ as a function of $\sqrt{s}$ in GeV. Data: 
brown triangles from Ref.~\cite{Baldenegro:2024vgg},
blue circles from Ref.~\cite{TOTEM:2020zzr}. Theoretical curves (\ref{eq:ratioRbd})
as in Fig.~\ref{fig:total}.}%
\label{fig:Rbd}%
\end{figure}

In Fig.~\ref{fig:Rbd} we plot ratio ${\cal R}_{\rm bd}$ (\ref{eq:ratioRbd}) for two parametrizations (\ref{eq:parPDG}) and (\ref{eq:parDL})
with $c_0\simeq 0.03$. One can see again that the ratio data are well reproduced by both parametrizations.

Let us finally discuss the energy dependence of the total elastic cross-section (\ref{eq:Wpower}) as compared to $\sigma_{\rm tot}$.
Recall that the modification factor (\ref{eq:elvstot}) is equal to $1+c_1\rho^2(s)$. 
At the ISR, the parameter $\rho$ is very small, and despite the fact that it increases rapidly with energy, its influence
on  $\sigma_{\rm el}$ is negligible.\footnote{To be more quantitative we need to compute $c_1$.} 
At the LHC the $\rho$ parameter is slightly larger, however it is almost energy independent,
(see Fig.~\ref{fig:total}) and therefore does not influence the energy dependence of  $\sigma_{\rm el}$ as well.
Note, however, that formula (\ref{eq:sigeltot}) assumes that GS 
holds {\sl everywhere} in $t$. This is certainly not true outside
the dip\,--\,bump region, although at the ISR GS surprisingly holds even for very small $t$ \cite{Baldenegro:2024vgg}. 
Hence, the fact that GS does not describe well  the energy dependence of the total elastic cross section at the LHC should be attributed to
the GS violation outside the dip\,--\,bump region, most importantly at small $t$. 

\section{Summary}

We have explored the property  that bump to dip {\em positions} ratio of the elastic $pp$ cross section is constant
over the wide energy range from the ISR to the LHC, implying that the cross-section depends on the scaling variable $\tau\sim \sigma_{\rm tot} |t|$,
which aligns bump and dip {\em positions} at all energies. By imposing the crossing symmetry and the Taylor expansion around the
real part of $-is$, allowed us to identify the imaginary and real parts of $\tilde{T}_{\rm el}(s,t)$.
This in turn allowed us to calculate the $\rho$ parameter
and the ${\cal R}_{\rm bd}$ ratio (\ref{eq:Rbddef}). The agreement with data is very good.
This means that the main properties of total and 
differential cross sections at all energies can be explained from a simple and intuitive picture based on GS.
However, this approach is only approximate, it fails to reproduce the energy dependence
of $\sigma_{\rm el}$ at the LHC due to the violation of GS at small $t$.

\end{document}